# Harmonic-Recycling Rectification Based on Novel Compact Dual-Band Resonator

Pengde Wu, Hao Wu, Yi-Dan Chen, Zhi Hua Ren, Yuhua Cheng, and Changjun Liu, *Senior Member, IEEE*

*Abstract*—Harmonic generation during RF-DC conversion causes performance degradation of a microwave rectifying circuit. To suppress and recycle the harmonic power, this letter proposes a novel compact dual-band resonator (DBR) based on a microstrip coupled transmission line. It presents open-circuits at the second and third harmonic frequencies, which effectively block the higher-order harmonic for power recycling. The conventional input cascading filters for harmonic rejection can be eliminated, simplifying the circuit topology and reducing loss. Theoretical analyses were carried out and corresponding equations were formulated for the proposed DBR. For validation, two rectifying circuits with/without the DBR operating at 2.2 GHz were fabricated and tested. Using the proposed DBR at 10 dBm RF power, the suppression of the second and third harmonic powers is enhanced by 18.4 dB and 7.6 dB, respectively. Besides, an improvement of RF-DC power efficiency (PCE) was observed; specifically, PCE reached 73.2% at 10 dBm compared to 71.6% obtained from an equivalent rectifier.

## I. INTRODUCTION

MICROWAVE rectifying circuits, essential components in wireless power transmission (WPT) systems, play a crucial role in converting RF power to DC power [1], [2], [3]. The overall efficiency of wireless power transfer (WPT) systems heavily relies on the power conversion efficiency (PCE) of the rectifying circuit [4], [5], [6], [7]. Therefore, the primary requirement of a rectifying circuit is high RF–DC power conversion efficiency (PCE), and miniaturization is another requirement for system integration in the antenna array.

In a rectifying circuit, the diode generates frequency harmonics from the incoming power, thereby reducing the proportion of energy converted to direct current and resulting in unwanted harmonic emissions. Moreover, as the incident radio frequency (RF) energy increases, the energy lost to harmonics further increases [8], [9]. Consequently, recycling rectification of the harmonic power is an effective method to boost the PCE of a rectifying circuit.

This work was supported by NSFC under Grant 62101366 and Grant U22A2015 . (Corresponding authors: Yuhua Chen, Changjun Liu)

P. Wu, H. Wu, and Y. Cheng are all with the Key Laboratory of Micro-Nano Sensing and IoT of Wenzhou, Wenzhou Institute of Hangzhou Dianzi University, Wenzhou 325038, China, and they are also with the School of Electronics and Information, Hangzhou Dianzi University, Hangzhou 311101, China (e-mail: chengyh@hdu.edu.cn).

Y.D. Chen is with Georgia Institute of Technology, School of Material Science and Engineering, College of Engineering, North Ave NW, Atlanta, GA 30332, United States.

Z.H. Ren is with the School of Biomedical Engineering & State Key Laboratory of Advanced Medical Materials and Devices, ShanghaiTech University, Shanghai, 201210, China.

C. Liu is with the School of Electronics and Information Engineering, Sichuan University, Chengdu, 610064, China. (e-mail: cjliu@ieee.org).

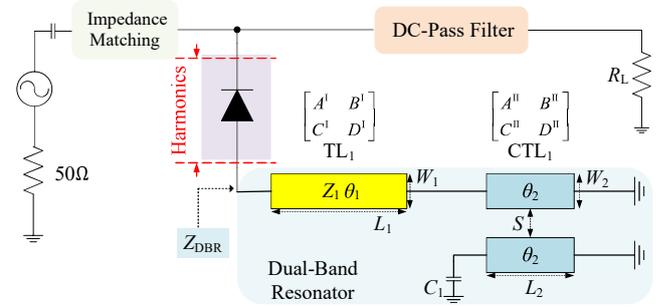

Fig. 1. A microwave rectifying circuit using a DBR for harmonic recycling.

Filtering of the harmonics at both the input and output has been investigated, as shown in [10] and [11], mainly to reduce re-radiated harmonic power. Usually, filters, e.g., band-stop [12], or low-pass filter [13] are applied between the diodes and the input/output of the microwave rectifier to reflect harmonics produced during rectifying, and the reflected harmonics go back to the diode where they can be converted to DC. It is noticed that the filters at the input raise a concern about the insertion loss and circuit area.

Class-F loads and harmonically terminated techniques are alternatives to the low-pass or dc-pass of rectifiers [14], it has good potential applications for high-frequency microwave rectifiers. Diode rectifiers using Class-C loads [15], [16] and a 5.8-GHz charge pump rectifier with Class-F loads [17] demonstrated an improvement in PCE. In reality, the optimum impedance for harmonic-terminated operation is difficult to achieve as the package of the device has parasitics that move the harmonic impedances.

Recently, a competitive solution has emerged in the form of a compact serial bandstop structure. This innovative approach not only effectively blocks radiation of harmonics but also compensates for the diode capacitive impedance at the fundamental frequency [7], [18]. In [7], a short-ended eighth-wavelength microstrip transmission line was employed to impede the second harmonic and enhance power recycling. However, solely the second harmonic can be recycled.

In this letter, we propose a novel harmonic suppression structure using a dual-band resonator (DBR) to efficiently recycle both the second and third harmonic power with a compact circuit topology. The DBR is based on our previous work [19], where we found that a microstrip line in serial with a coupled pair of microstrip-lines could provide two poles (open-circuits) for impeding harmonic radiation. In this letter, we further investigate this structure as a DBR to demonstrate its ability to block the second and third harmonics for enhanced power recycling compared to [7]. With the proposed DBR, a simple and compact circuit structure was realized, resulting in a decrease in insertion loss and an improvement of PCE.

## II. Principle and Design Method

### A. Principle

As shown in Fig. 1, owing to the high impedance provided by the harmonic suppression structure, high-order harmonics are confined to the diode without returning to the input port, thus the conversion efficiency of a rectifying circuit can be improved. However, In this design, the harmonic suppression structure in series with the diode has been realized through a DBR which is placed between the diode and the ground plane. A DBR in series with the diode should have the following functions:

1) Presenting a short circuit at DC, which provides a DC path for the rectifying circuit output.
2) Compensating the capacitive impedance of the diode, which facilitates the impedance matching circuit of the rectifying circuit.
3) Exhibiting high reflections at harmonic frequencies $nf_0 (n = 2, 3)$, which blocks them from returning the input port.

Fig. 1 shows a rectifying circuit operating with a DBR for harmonic recycling, it has three parts: 1) a quarter-wave transformer, which may be needed for the impedance matching at $f_0$; 2) a DC-pass filter to prevent RF signals from entering the load; 3) a DBR, employed to recycle the second and third harmonic power through rectifying or mixing with other frequencies.

### B. The proposed DBR

Based on the above analysis, the input impedance of the proposed DBR for harmonic recycling can be expressed as

$$\begin{cases} Z_{\text{DBR}}\ @\text{DC} = 0 \\ Z_{\text{DBR}}\ @f_0 = jX_L \\ Z_{\text{DBR}}\ @2f_0 = \infty \\ Z_{\text{DBR}}\ @3f_0 = \infty \end{cases} \quad (1)$$

A practical microstrip band-stop structure to block the third harmonics is a short-ended $\lambda/12$ transmission line. Inspired by this idea, a small section of a short-ended $\lambda/12$ transmission line is modified with a coupled transmission line ($\text{CTL}_1$), which is shown Fig. 1. Due to the odd and even modes of the $\text{CTL}_1$, besides the third harmonic, another resonant frequency at the second harmonic can be simultaneously obtained.

Fig. 1 shows the schematic of the proposed DBR. The impedance and electrical length of $\text{TL}_1$ are $Z_1$ and $\theta_1$, respectively, thus its ABCD matrix [20] is

$$\begin{bmatrix} A^{\text{I}} & B^{\text{I}} \\ C^{\text{I}} & D^{\text{I}} \end{bmatrix} = \begin{bmatrix} \cos\theta_1 & jZ_1\sin\theta_1 \\ j\frac{\sin\theta_1}{Z_1} & \cos\theta_1 \end{bmatrix}. \quad (2)$$

Width $W_2$ and gap $S$ determine the even and odd-mode impedance of $\text{CTL}_1$, $Z_{0e}/Z_{0d}$. The ABCD matrix of $\text{CTL}_1$ can be expressed as

$$\begin{bmatrix} A^{\text{II}} & B^{\text{II}} \\ C^{\text{II}} & D^{\text{II}} \end{bmatrix} = \begin{bmatrix} \frac{Z_{0e}+Z_{0d}}{Z_{0e}-Z_{0d}} & j\frac{2\tan\theta_2}{(Z_{0e}-Z_{0d})} \\ -j\frac{2\cot\theta_2}{Z_{0e}-Z_{0d}} & \frac{Z_{0e}+Z_{0d}}{Z_{0e}-Z_{0d}} \end{bmatrix}. \quad (3)$$

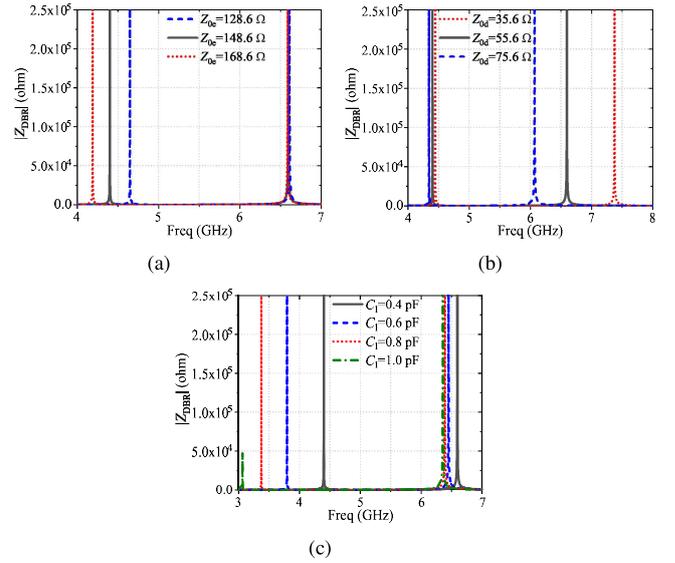

Fig. 2. Calculated $|Z_{\text{DBR}}|$ versus RF frequency with perturbation of (a) $Z_{0e}$. (b) $Z_{0d}$. (c) $C_1$.

Cascading the ABCD matrices in Equations (2) and (3) leads to

$$\begin{bmatrix} A^{\text{III}} & B^{\text{III}} \\ C^{\text{III}} & D^{\text{III}} \end{bmatrix} = \begin{bmatrix} A^{\text{I}} & B^{\text{I}} \\ C^{\text{I}} & D^{\text{I}} \end{bmatrix} \times \begin{bmatrix} A^{\text{II}} & B^{\text{II}} \\ C^{\text{II}} & D^{\text{II}} \end{bmatrix}. \quad (4)$$

Finally, the input impedance of the proposed DBR, with one port terminated in $C_1$, is

$$Z_{\text{DBR}} = \frac{A^{\text{III}}Z_C + B^{\text{III}}}{C^{\text{III}}Z_C + D^{\text{III}}} \quad (5)$$

where $Z_C$ is the impedance of capacitor $C_1$.

To verify the input impedance calculation and the performance of the proposed DBR, numerical experiments were carried out for a DBR operating at a fundamental frequency of 2.2 GHz. Considering the fabrication precision of PCB manufacture, the default width $W_2$ and gap $S$ of $\text{CTL}_1$ are 0.3 mm and 0.2 mm, respectively. Since $\text{TL}_1$ is in series with $\text{CTL}_1$, its width $W_1$ stays the same as $W_2$. Thus, the initial parameters of DBR are $Z_1 = 107.6\ \Omega$, $Z_{0e} = 148.6\ \Omega$, and $Z_{0d} = 55.6\ \Omega$ with an RF laminate RO4350B of 0.76-mm thickness. When $C_1$ is 0.4 pF, $\theta_1$ and $\theta_2$ are calculated as 15.9° and 18.6°, respectively. Substituting the above parameters into (5) for calculation, Fig. 2 shows the calculated $|Z_{\text{DBR}}|$ with perturbation of $Z_{0e}$, $Z_{0d}$, and $C_1$, respectively.

The frequency response of the DBR was studied through sweeping each parameter individually. As shown in Fig. 2(a), the first pole of $|Z_{\text{DBR}}|$ shifts to the left side dynamically with an increase of $Z_{0e}$, whereas the second pole of $|Z_{\text{DBR}}|$ moves to the right side rapidly with a decrease of $Z_{0d}$ (see Fig. 2(b)). Regarding the influence of capacitor $C_1$, Fig. 2(c) illustrates the shift in $|Z_{\text{DBR}}|$ as $C_1$ increases towards infinity (effectively a short circuit), with the other parameters remain the same as those in Fig. 2(a). It is observed the second pole of $|Z_{\text{DBR}}|$ stays the same place because $C_1$ is equivalent to a short circuit at a relatively high frequency, whereas the first pole shifts towards lower frequencies as $C_1$ increases. As a result, the first pole can be tuned through $C_1$. However, a large $C_1$ may place the first pole very close to $f_0$, which should be avoided.

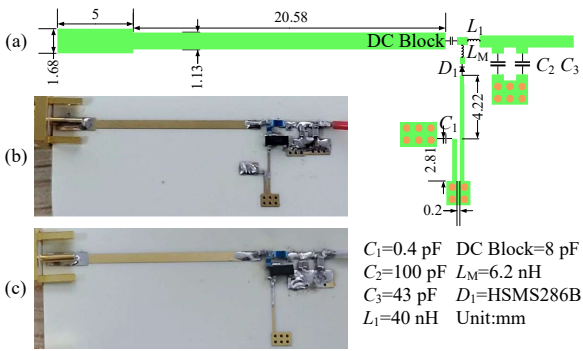

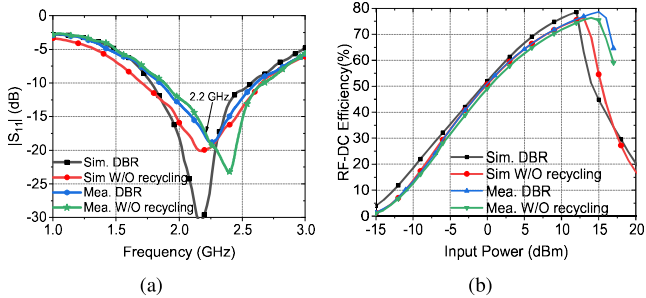

Fig. 3. (a) Circuit layout with design parameters for the proposed DBR. (b) A rectifying circuit with the proposed DBR. (c) A rectifying circuit without harmonic recycling for performance comparison.

Fig. 4. (a) Simulated and measured $|S_{11}|$ versus RF frequency at 10-dBm input power and (b) PCE versus RF input power.

## III. IMPLEMENTATION AND MEASUREMENT

A rectifying circuit circuit operating at $f_0 = 2.2$ GHz was simulated in Advanced Design System (ADS, keysight) with physical dimensions indicated in Fig. 3, and the RO4350B ($\epsilon_r = 3.66$ and $\tan\delta = 0.002$) was used. $L_M$ is a chip inductor for compensating the capacitive impedance of the diode. Fig. 3 shows a photograph of the fabricated rectifying circuit. For performance comparison, a rectifying circuit without harmonic recycling was also fabricated.

The measured reflection coefficient $|S_{11}|$ at the input power of 0 and 10 dBm are shown in Fig. 4(a). For the rectifying circuit with the DBR, $|S_{11}|$ curve for 10 dBm is well below -10 dB for frequencies ranging from 1.85 to 2.6 GHz. It is observed that $|S_{11}|$ for rectifying circuits with/without the DBR are almost the same.

Fig. 4(b) depicts the simulated measured PCE versus input power at 2.2 GHz and 400 Ω DC load. As observed at 10 dBm, the measured PCE of the rectifying circuits with/without the proposed DBR are 73.2% and 71.6% respectively, while simulated PCE for both rectifiers are 76.2% and 73.4%, respectively. Because the measured $|S_{11}|$ versus input power are not same for both rectifiers, the improvement of PCE by the measurement is lower than that of the simulation. Compared to the reference rectifier, the rectifying circuit using the proposed DBR shows an improvement in PCE at high input power levels (0 to 14 dBm).

The measured power levels of the second and third harmonic are depicted in Fig. 5(a) where the rectifying circuit with DBR presents an obvious low point at 2.2 GHz. Specifically, at 10 dBm, the suppression of the second harmonic is enhanced from -6.7 dBm to -25.1 dBm by the proposed DBR (an improvement of 18.4 dB). Meanwhile, as shown in Fig. 5(b), the rectifying circuit with DBR demonstrates better control over the third harmonic, showing an improvement of 7.6 dB by comparison. The measured harmonic power versus frequency validates the effectiveness of harmonic control through the proposed DBR.

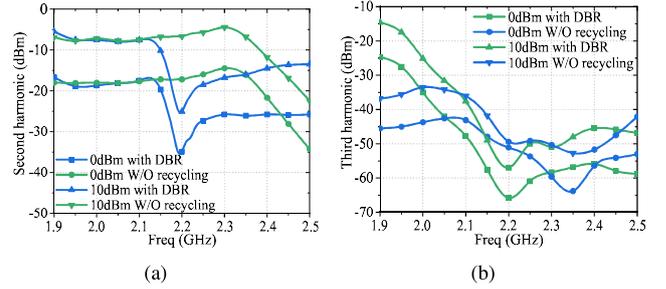

Fig. 5. Measured harmonic power versus RF frequency at 0- and 10-dBm input power. (a) Second harmonic. (b) Third harmonic.

TABLE I
PERFORMANCE COMPARISON OF HARMONIC-RECYCLING RECTIFIERS

| Ref. | [7] | [18] | [21] | This work |
|---|---|---|---|---|
| Freq (GHz) | 2.45 | 2.45 | 2.6 | 2.2 |
| PCE (%) | 80.9 | 80.2 | 47 | 78.6 |
| Pow.(dBm) | 20 | 25 | 0 | 15 |
| Diode | HSMS-282 | HSMS-282 | HSMS-2850,2860 | HSMS-286B |
| Size (mm) | 16 × 18 | 20 × 5 | 37 × 31 | 34 × 11 |
| Size ($\lambda_g^2$) | 0.051 | 0.018 | 0.219 | 0.073 |
| Technology | TL. $\lambda/8$ short-ended | TL. $\lambda/8$ short-ended | Another rectifier cell | Dual-band resonator |
| Harmonic Supp. (dB) | N.A | N.A | N.A | 18.4@$2f_0$ 7.6@$3f_0$ |
| Diode Num. | 1 | 2 | 2 | 1 |

Table I shows a comparison of the performances between the proposed rectifying circuit and those reported in the literature targeting harmonic recycling. As shown, the rectifying circuit using the proposed DBR has demonstrated the effectiveness of harmonic control over the second and third harmonics, while harmonic suppressions in other references were not reported. Meanwhile, only one diode is used, which helps to minimize the losses associated with the diode turn-on voltage.

## IV. CONCLUSION

A novel compact dual-band resonator (DBR) has been developed for recycling the second and third harmonic power in a microwave rectifying circuit. It functions as both a filter and an inductor for impedance matching, the conventional input cascading band-pass or low-pass filters for harmonic rejection can be eliminated, leading to a simple circuit topology with low loss. Compared to a reference rectifier, the proposed DBR demonstrates better control over the second and third harmonic, showing an improvement of 18.4 dB and 7.6 dB respectively. Besides, improved PCEs at high input power levels were also observed.

This article has been accepted for publication in IEEE Microwave and Wireless Technology Letters. This is the author's version which has not been fully edited and content may change prior to final publication. Citation information: DOI 10.1109/LMWT.2025.3528521IEEE MICROWAVE AND WIRELESS TECHNOLOGY LETTERS 4## References

[1] J. Huang, Y. Zhou, Z. Ning, and H. Gharavi, "Wireless power transfer and energy harvesting: Current status and future prospects," *IEEE wireless communications*, vol. 26, no. 4, pp. 163–169, 2019.

[2] S.-P. Gao, J.-H. Ou, X. Zhang, and Y. Guo, "Scavenging microwave wireless power: A unified model, rectenna design automation, and cutting-edge techniques," *Engineering*, 2023.

[3] P. Wu, S. Y. Huang, W. Zhou, W. Yu, Z. Liu, X. Chen, and C. Liu, "Compact high-efficiency broadband rectifier with multi-stage-transmission-line matching," *IEEE Transactions on Circuits and Systems II: Express Briefs*, vol. 66, no. 8, pp. 1316–1320, 2019.

[4] S. Ladan, A. B. Guntupalli, and K. Wu, "A high-efficiency 24 GHz rectenna development towards millimeter-wave energy harvesting and wireless power transmission," *IEEE Transactions on Circuits and Systems I: Regular Papers*, vol. 61, no. 12, pp. 3358–3366, 2014.

[5] D. Mishra, S. De, and K. R. Chowdhury, "Charging time characterization for wireless RF energy transfer," *IEEE Transactions on Circuits and Systems II: Express Briefs*, vol. 62, no. 4, pp. 362–366, 2015.

[6] M.-D. Wei, Y.-T. Chang, D. Wang, C.-H. Tseng, and R. Negra, "Balanced RF rectifier for energy recovery with minimized input impedance variation," *IEEE Transactions on Microwave Theory and Techniques*, vol. 65, no. 5, pp. 1598–1604, 2017.

[7] C. Liu, F. Tan, H. Zhang, and Q. He, "A novel single-diode microwave rectifier with a series band-stop structure," *IEEE transactions on microwave theory and techniques*, vol. 65, no. 2, pp. 600–606, 2017.

[8] S. Ladan and K. Wu, "Nonlinear modeling and harmonic recycling of millimeter-wave rectifier circuit," *IEEE Transactions on Microwave Theory and Techniques*, vol. 63, no. 3, pp. 937–944, 2015.

[9] T.-W. Yoo and K. Chang, "Theoretical and experimental development of 10 and 35 GHz rectennas," *IEEE Transactions on Microwave Theory and Techniques*, vol. 40, no. 6, pp. 1259–1266, 1992.

[10] S. Imai, S. Tamaru, K. Fujimori, M. Sanagi, and S. Nogi, "Efficiency and harmonics generation in microwave to DC conversion circuits of half-wave and full-wave rectifier types," in *2011 IEEE MTT-S International Microwave Workshop Series on Innovative Wireless Power Transmission: Technologies, Systems, and Applications*. IEEE, 2011, pp. 15–18.

[11] J.-Y. Park, S.-M. Han, and Itoh, "A rectenna design with harmonic-rejecting circular-sector antenna," *IEEE Antennas and Wireless Propagation Letters*, vol. 3, pp. 52–54, 2004.

[12] H. Lin, X. Chen, Z. He, Y. Xiao, W. Che, and C. Liu, "Wide input power range X-band rectifier with dynamic capacitive self-compensation," *IEEE Microwave and Wireless Components Letters*, vol. 31, no. 5, pp. 525–528, 2021.

[13] J. O. McSpadden, L. Fan, and K. Chang, "Design and experiments of a high-conversion-efficiency 5.8-GHz rectenna," *IEEE Transactions on Microwave Theory and techniques*, vol. 46, no. 12, pp. 2053–2060, 1998.

[14] J. Guo, H. Zhang, and X. Zhu, "Theoretical analysis of RF-DC conversion efficiency for class-F rectifiers," *IEEE transactions on microwave theory and techniques*, vol. 62, no. 4, pp. 977–985, 2014.

[15] M. Roberg, E. Falkenstein, and Z. Popović, "High-efficiency harmonically-terminated rectifier for wireless powering applications," in *2012 IEEE/MTT-S International Microwave Symposium Digest*. IEEE, 2012, pp. 1–3.

[16] Z. He, H. Lin, and C. Liu, "A novel class-C rectifier with high efficiency for wireless power transmission," *IEEE Microwave and Wireless Components Letters*, vol. 30, no. 12, pp. 1197–1200, 2020.

[17] C. Wang, N. Shinohara, and T. Mitani, "Study on 5.8-GHz single-stage charge pump rectifier for internal wireless system of satellite," *IEEE Transactions on Microwave Theory and Techniques*, vol. 65, no. 4, pp. 1058–1065, 2017.

[18] Z. He, H. Lin, H. Zhu, and C. Liu, "A compact high-efficiency rectifier with a simple harmonic suppression structure," *IEEE Microwave and Wireless Components Letters*, vol. 30, no. 12, pp. 1177–1180, 2020.

[19] P. Wu, S.-P. Gao, Y.-D. Chen, Z. H. Ren, P. Yu, and Y. Guo, "Harmonic-based integrated rectifier–transmitter for uncompromised harvesting and low-power uplink," *IEEE Transactions on Microwave Theory and Techniques*, vol. 71, no. 2, pp. 870–880, 2023.

[20] D. M. Pozar, *Microwave engineering: theory and techniques*. John wiley & sons, 2021.

[21] T. Ngo and Y.-X. Guo, "Harmonic-recycling rectifier for high-efficiency far-field wireless power transfer," *IEEE Transactions on Circuits and Systems II: Express Briefs*, vol. 67, no. 4, pp. 770–774, 2020.© 2025 IEEE. Personal use is permitted, but republication/redistribution requires IEEE permission. See https://www.ieee.org/publications/rights/index.html for more information.